\documentclass[10pt,prl,twocolumn]{revtex4}
\usepackage{rotating}
\usepackage{graphicx}
\usepackage{amsmath,amssymb,amsfonts}

\def\Ca{{C$_\alpha$}}

\def\Ca{{C$_\alpha$}}

\begin{document}
\title{Internal protein dynamics shifts the distance to the mechanical
  transition state}
\author{Daniel K. West}
\affiliation{School of Physics \& Astronomy}
\affiliation{School of Biochemistry \& Microbiology, 
University of Leeds, Leeds LS2 9JT, United Kingdom}
\author{Emanuele Paci}
\affiliation{School of Physics \& Astronomy}
\affiliation{Astbury Centre for Structural Biology,
University of Leeds, Leeds LS2 9JT, United Kingdom}
\author{Peter D. Olmsted}
\email{p.d.olmsted@leeds.ac.uk}
\affiliation{School of Physics \& Astronomy}
\affiliation{Astbury Centre for Structural Biology,
University of Leeds, Leeds LS2 9JT, United Kingdom}
\begin{abstract}
  Mechanical unfolding of polyproteins by force spectroscopy provides
  valuable insight into their free energy landscapes. Most
  phenomenological models of the unfolding process are two-state
  and/or one dimensional, with the details of the protein and its
  dynamics often subsumed into a zero-force unfolding rate and a
  single distance $x_u^{1\textrm{D}}$ to the transition state.  We
  consider the entire phase space of a model protein under a constant
  force, and show that the distance $x_u^{1\textrm{D}}$ contains a
  sizeable contribution from exploring the full multidimensional
  energy landscape.  Proteins with more degrees of freedom are
  expected to have larger values for $x_u^{1\textrm{D}}$. We show that
  externally attached flexible linkers also contribute to the measured
  unfolding characteristics.
\end{abstract}
\maketitle

Atomic force microscopy or optical tweezers are now routinely used to
study the mechanical properties of proteins \cite{rief97,rief98a}.  An
important issues is the unfolding behavior of folded domains,
including the strength, and the dependence on fold topology
\cite{west06} and secondary structure \cite{ortiz05,west06}.  The
simplest description of unfolding treats the unfolding domain as
moving in a one dimensional potential $G(x)$, where the reaction
coordinate $x$ is assumed to be directly coupled to the applied force.
In perhaps the simplest approximation, the rate of unfolding $k_u(F)$
of a two-state (native and denatured) protein under force can then be
calculated using Bell's formula based on Kramers' relation for escape
from a well \cite{bell78,rief97}:
\begin{equation}\label{eqn:Bell}
k_u(F)\simeq k_u^0\exp\left(\frac{Fx_u^{1\textrm{D}}}{k_BT}\right),
\end{equation}
where $k_B$ is Boltzmann's constant, $T$ is the absolute temperature,
$x_u^{1\textrm{D}}$ is the transition state displacement along the
force projection, and $k_u^0$ is the unfolding rate constant at zero
force. This result can be used to calculate the distributions of
unfolding times (for applied force) or unfolding forces (for applied
pulling speed), and has been applied to many different proteins
\cite{fernandez04,schlierf04,oberhauser01,brockwell05,brockwell02}.
This simple result has been corrected to use the entire shape of
$G(x)$ rather than just the barrier height and displacement
$x_u^{\mathrm{1D}}$ and incorporate the cantilever compliance
\cite{hummer03,schlierf06}, to relax the diffusive limit to faster
speeds \cite{dudko2006}, and to allow for multiple pathways or states
\cite{Bartolo02,williams03}. There has been extensive work on the
utility of a the $x_u$ and the simple 1D picture, linear force
dependence, etc \cite{li04,kirmizialtin05} \textbf{[More work to do
  here!!!]}

\begin{figure}[hb]
\begin{center}
\includegraphics[width=0.99\linewidth]{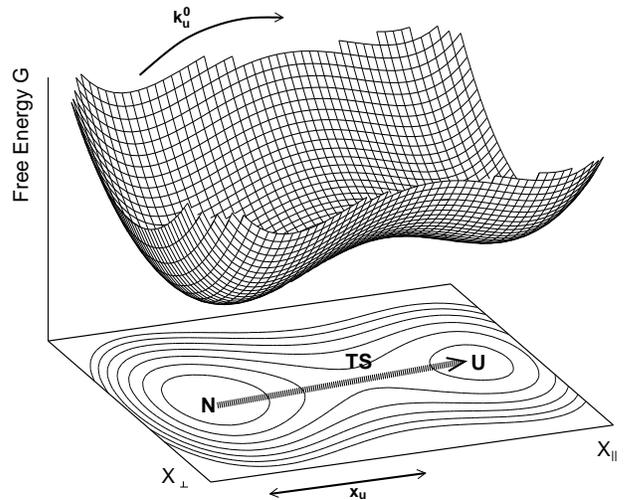}
\caption{The free energy surface of a two-state system, with 
  native (N) and unfolded (U) minima. An external force $F$ parallel
  to $\textrm{X}_{\parallel}$ lowers the barrier to unfolding by
  $Fx_u$.}
\label{fig:MD_free_energy}
\end{center}
\end{figure}
However, the assumption of a one dimensional reaction coordinate
grossly simplifies physical reality.  The unfolding rate depends
dramatically on pulling direction
\cite{carrion-vazquez03,brockwell03}, and hence on the
multidimensional nature of the free energy landscape.  Moreover, the
one dimensional parameters have no satisfactory physical
interpretation: $x_u^{1\textrm{D}}$ is defined along the pulling
direction, while the actual unfolding takes place along an unknown
reaction coordinate(s), presumably involving a few key amino acids.
While $x_u^{\textrm{1D}}\simeq0.2\,\mathrm{nm}$ is the right order of
magnitude for hydrogen bonds, the explicit connection with molecular
configurations remains unclear.  In this Letter we explore the role of
the multidimensional energy landscape in determining the effective 1D
unfolding parameters.  The key physical ingredient is that an applied
force perturbs fluctuations transverse to the forcing direction,
because of the highly anharmonic nature of angular bonds.
Specifically, force restricts a protein's conformational search among
the dihedral states of the polypeptide backbone.  We calculate this
using molecular dynamics (MD) simulations of a simple off-lattice G\=o
model \cite{karanicolas02,west06b} for a topologically simple protein,
and show that this restriction leads to a sizeable contribution to
$x_u^{1\textrm{D}}$.

\textit{Escape from a multidimensional energy landscape---} The rate
of escape from an ${\cal N}$-dimensional energy landscape under an
applied force $F$ (Fig.~\ref{fig:MD_free_energy}) is given by
\cite{haenggi90}
\begin{align}
  k_u(F)&=\Gamma (F)\exp\left(\frac{Fx_u}{k_BT}\right), \label{eqn:Kramer_MD}\\
  \Gamma (F)&=\frac{\sqrt{|G^{''RC}_{TS}|}}{2\pi
    \gamma}\frac{\prod_{k=1}^{{\cal N}}\sqrt{G^{''k}_N(F)}
    }{\prod_{k=1}^{{\cal N}-1}\sqrt{G^{''k}_{TS}}
    }\exp\left(\frac{-\Delta
      G_{TS-N}}{k_BT}\right),\label{eqn:Kramer_Gamma}
\end{align}
where $x_u$ is the distance to the transition state, $\gamma$ is a
friction coefficient, and $\Delta G_{TS-N}$ is the height of the free
energy barrier relative to the native basin.  $G^{''RC}_{TS}$ is the
curvature in the unstable direction at the transition state,
$G^{''k}_{TS}$ are the ${\cal N}-1$ stable curvatures at the
transition state, and $G^{''k}_N(F)$ are the ${\cal N}$ positive
curvatures about the native basin under an applied force.  If the
transition state is sharp {\it i.e.,} $|G^{''RC}_{C}| \gg {F}/{x_u}$,
the curvatures $G^{''}_{TS}$ at the transition state, as well as
$x_u$, are approximately independent of force.  A more physical
representation of the attempt frequency $\Gamma(F)$ follows by
relating a curvature $G^{''k}$ to the associated fluctuations by
$\langle \delta x_k^2\rangle=k_BT/G^{''k}$. This yields
\begin{equation}
\Gamma(F)=\frac{k_BT}{2\pi
  \gamma l_{TS}^{RC}}\frac{V_{TS}}{V_N(F)} \exp\left(\frac{-\Delta
    G_{TS-N}}{k_BT}\right),\label{eqn:Kramer_Vol} 
\end{equation}
where $l_{TS}^{RC}$ is the width of the transition state along the
unfolding reaction coordinate, $V_{TS}$ is the volume of phase space
available for fluctuations at the transition state and $V_N(F)$ is the
corresponding volume in the native basin.
  
In one dimension the weak force dependence of the prefactor $\Gamma
(F)$ can be safely ignored \cite{evans97}.  A weak perturbation of the
native basin volume $V_N(F)$ leads to
\begin{equation} \label{eqn:xu_renorm}
k_u(F)\simeq k_u^0 \exp\left(\frac{F\left[x_u+\lambda k_BT\right]}{k_BT}\right),
\end{equation}
where $\lambda=-\left.{\partial[\ln V(F)]}/{\partial F}\right|_{F=0}$.
Thus, any change in the volume of the native basin with force will
shift $x_u^{1\textrm{D}}$ in the equivalent one dimensional model,
\begin{align}
  x_u^{1\textrm{D}}&= x_u+\lambda k_BT\equiv x_u + \delta x_u, \label{eq:deltaxu}
\end{align}
where $\delta x_u$ is a \textit{dynamical}, or entropic, contribution
to the transition state displacement.  If the volumes associated with
different degrees of freedom randomly increased or decreased with an
applied force, there would be little effect. However, we expect the
volume of most perturbed degrees of freedom to decrease under an
applied force, so that $\delta x_u$ is proportional to the number of
perturbed degrees of freedom.  For a force-independent volume we
recover Eq.~\ref{eqn:Bell}, with $\delta x_u=0$.

\begin{figure}[hb]
\begin{center}
\includegraphics[width=0.95\linewidth]{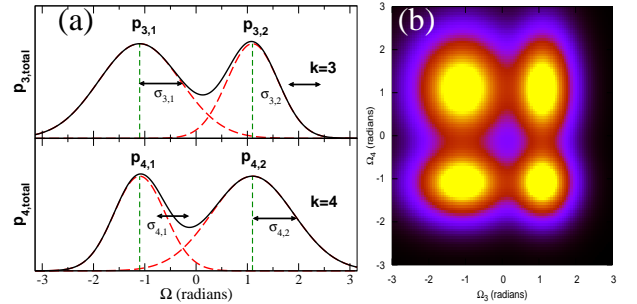}
\caption{The  phase space of an oligomer with $4$
  dihedral angles: two of these are unimodal ($k=1,2$, not shown) and
  two are bimodal ($k=3,4$).  (a) The probability distribution
  function $p_{k,\mathrm{total}}$ for each bi-modal dihedral angle
  (black solid line) can be resolved into separate distributions
  $p_{k,n}$ (red dashed lines) about well defined averages (green
  vertical dotted lines). There are four possible structures
  corresponding to fluctuations around
  \{$\bar{\Omega}_{3,1}$,$\bar{\Omega}_{4,1}$\},\{$\bar{\Omega}_{3,1}$,$\bar{\Omega}_{4,2}$\},
  \{$\bar{\Omega}_{3,2}$,$\bar{\Omega}_{4,1}$\} and
  \{$\bar{\Omega}_{3,2}$,$\bar{\Omega}_{4,2}$\} (b) The phase space
  projected onto \{$\Omega_3$,$\Omega_4$\}.}
\label{fig:phasespace}
\end{center}
\end{figure}

\textit{Calculation of phase space volumes---} Phase space fluctuation
volumes were calculated from MD simulation trajectories.  MD was
performed for protein L (PDB reference: \textsc{1HZ6} \cite{oneill01})
using the \Ca\ G\=o model of
Refs.~\cite{karanicolas02,karanicolas03a}.  The simulation protocol is
described in detail in \cite{west06}.  A real protein does not
fluctuate about a single well-defined average structure; because
dihedral angles typically access discrete values, the accessible
states are fluctuations about many well defined structures, or
\textit{nodes} in phase space.  Fig.~\ref{fig:phasespace} shows the
phase space explored by a tetramer with 2 unimodal and 2 bimodal
dihedral distributions.

The total unfolding rate $k_u^{\mathrm{tot}}(F)$ is the weighted sum
of the escape rates $k^{\beta}_{u}(F)$ from all nodes $\beta=1\ldots M$
(assuming $\Delta G_{TS-N}$ and $x_u$ are the same for all nodes);
\begin{align}\label{eqn:rates}
  k^{\mathrm{tot}}_{u}(F) &=\frac{k_BT}{2\pi\gamma l_{TS}^{RC}}
  \frac{V_{TS}}{V_N^{\mathrm{eff}}(F)}
  \exp{\left[-\frac{\left(\Delta G_{\scriptscriptstyle TS-N} -
          Fx_u\right)}{k_BT}\right]} ,\\
  \frac{1}{V_N^{\mathrm{eff}}(F)}&= \sum_{\beta =1}^M
  \frac{P^{\beta}(F)}{V_N^{\beta}(F)}, \label{eqn:volume}
\end{align}
where $V_N^{\beta}(F)$ is the volume and $P^{\beta}(F)$ the occupation
probability of node $\beta$. The quantity $V_N^{\mathrm{eff}}(F)$ is
the {\it effective} phase space volume of the native basin.

The occupation probability $P^{\beta}(F)$ of each node is 
\begin{equation}\label{eqn:configprob}
P^{\beta}(F)=\left<\prod_{k=1}^{N-3} \frac{p_{k,n_\beta} (\Omega_k(t))
    }{p_{k,\mathrm{total}}(\Omega_k(t))}\right>,
\end{equation}
where $\sum_{\beta=1}^{M}P^{\beta}(F)=1$ and $\langle\ldots\rangle$ is
the average over the MD trajectory.  A node is specified by a
particular set of occupancies of each dihedral angle
$\boldsymbol{\Omega} = (\Omega_1,\ldots\Omega_{N-3})$, where $N$ is
the number of atoms. Each term in the product is the normalised
probability that, in node $\beta$, a given dihedral angle $\Omega_k$
participates in its $n$th dihedral state (peak) (Fig.
\ref{fig:phasespace}a). For large numbers of nodes $M$ a mean field
approach, in which all nodes are assumed to be equally populated,
works well when states are sufficiently uncorrelated in time, as in
this case \cite{west06c}.

The volume $V_N^{\beta}(F)$ of fluctuations about each node $\beta$ is
given by $V_N^{\beta}(F)=\sqrt{\mathrm{det}\mathbf{C}^{\beta}}$, where
$\mathbf{C_{ij}}^{\beta}=\langle\delta
\mathbf{r}_i\delta\mathbf{r}_j\rangle_{\beta}$ is the covariance
matrix for fluctuations $\delta\mathbf{r}_i$ in each $C_{\alpha}$
position $\mathbf{r}_i$. Here the angle brackets denote an average
within node $\beta$. We calculate $\mathbf{C}^{\beta}$ by transforming
coordinates to bond lengths, bond angles, and dihedrals angles. We
ignore correlations between bond and dihedral angles, which is an
excellent approximation here \cite{west06c}.  Hence, the effective
volume of phase space is given by
\begin{equation}\label{eqn:vol_dih_ang}
\frac{1}{V_N^{\mathrm{eff}}(F)}\simeq\frac{1}{V_{\theta}(F)}
\sum_{\beta=1}^{M}\frac{P^{\beta}(F)}{V_{\Omega}^{\beta}(F)}. 
\end{equation}
where $V_{\theta}(F)$ and $V_{\Omega}^{\beta}(F)$ are the volumes of
phase space explored by the bond and dihedral angles respectively.

\begin{figure}[hb]
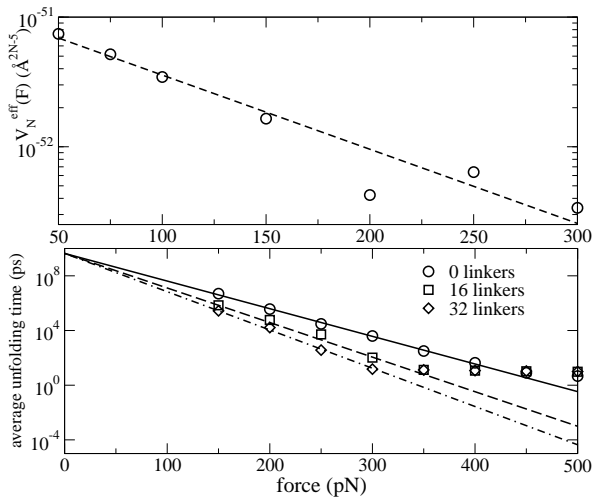

\begin{center}
\includegraphics[width=0.9\linewidth]{./1hz6_weight_meanfield_totB.eps}
\includegraphics[width=0.9\linewidth]{./linkB.eps}
\caption{(a) Effective phase space volume $V_N^{\mathrm{eff}}(F)$  for
  protein L ($\circ$). Dashed line is a fit to
  $\ln{V_N^{\mathrm{eff}}(F)} \simeq \ln{V_N^{\mathrm{eff}}(0)} -
  \delta x_u F/k_BT$, which yields the dynamic contribution to the
  transition state placement $\delta x_u=0.054\pm0.008\,\mathrm{nm}$
  (Eq.~\ref{eq:deltaxu}).  (b) Average unfolding times for protein L
  using MD at $T=300\,\mathrm{K}$, with $n_l$ attached
  glycine linkers ($\circ$: $n_l=0, x_u^{\rm 1D}=0.191\pm0.004
  ~\mathrm{nm}$, $\square$: $n_l=16, x_u^{\rm 1D}=0.241\pm0.004
  ~\mathrm{nm}$, $\diamond$: $n_l=32,x_u^{\rm
    1D}=0.267\pm0.004~\mathrm{nm}$). The linear fits yield $\log\tau =
  A - x_u^{1\mathrm{D}}F/(k_BT)$. Error bars are of order the symbol
  size.}
\label{fig:1hz6_tot}
\end{center}
\end{figure}

\textit{Results---} Fig.~\ref{fig:1hz6_tot}a shows the phase space
volume as a function of force calculated from MD simulations of
protein L.  The dynamic contribution to the transition state placement
is $\delta x_u=0.054\pm0.008\,\mathrm{nm}$.  The reduction of phase
space volume comes from (1) the narrowing of the dihedral
distributions, and (2) the reduction in the number of multi-modal
dihedral peaks. The latter effect dominates, since the loss of a
single dihedral peak immediately removes many nodes of phase space.
Simulations of the unfolding of the same protein L domain
(Fig.~\ref{fig:1hz6_tot}b) yield an effective 1D transition
displacement $x_u^{\mathrm{1D}}=0.191\pm0.004\,\mathrm{nm}$, from
measuring an exponential dependence of the unfolding time $\tau_u$ on
applied force, $\tau_u\sim e^{-Fx_u^{\mathrm{1D}}/(k_BT)}$, as
predicted from a single reaction coordinate description,
Eq.~(\ref{eqn:Bell}).  Hence we conclude that the bare transition
state position was $x_u=0.137\,\mathrm{nm}$, and the large shift of
$\delta x_u=0.054\pm0.008\,\mathrm{nm}$ is between 34\% and 45\%.

\textit{Linker Effects---} For convenience, protein domains are often
pulled with long linkers, or unfolded protein strands.  The linkers
\textit{also} fluctuate about discrete dihedral states when stretched.
The ``lumpiness'' of this phase space is irrelevant for weakly
stretched strands, but dominates the response for strongly stretched
strands.  Since force is coupled to the folded domain through the
linkers, the total available phase space is the product of protein and
linker phase spaces, and the measured $x_u^{\mathrm{1D}}$ depends on
the restriction of the linkers' phase space.

\begin{figure}[htb]
\begin{center}
\includegraphics[width=0.9\linewidth]{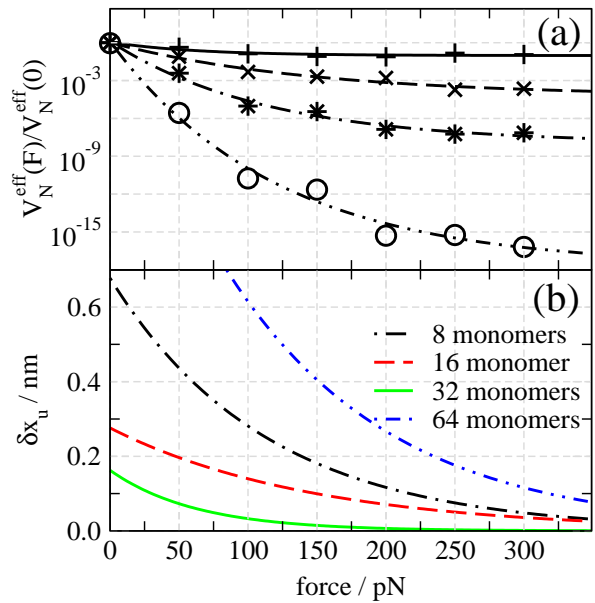}
\caption{(a) Volume of phase space calculated using the mean field
  method and (b) the dynamic contribution to the distance to the
  transition state $\delta x_u=-k_BT {\partial
    [\ln{V_N^{\mathrm{eff}}]}}/{\partial F}$, for strands of flexible
  glycine linkers. A constant force was applied for a total time of 1
  $\mu s$ at T=300K.}
\label{fig:partition}
\end{center}
\end{figure}

To test this, homogeneous linkers strands were constructed from a
dihedral potential based on glycine.  Fig.~\ref{fig:partition} shows
the normalised effective phase space volume
$V_N^{\mathrm{eff}}(F)/V_N^{\mathrm{eff}}(0)$ and the corresponding
$\delta x_u$ as a function of force for different number of atoms
$n_l$ per linker. The effect is greater for longer linkers, since more
nodes are available to remove.  We have ignored any force dependence
of the volume of the transition state $V_{TS}$
(Eq.~\ref{eqn:Kramer_Vol}).  Although we cannot easily characterize
the (unstable) transition state, we can compare the shifts $\delta
x_u$ measured directly from the unfolding times
(Fig.~\ref{fig:1hz6_tot}) with predictions from the phase space
volumes (Fig.~\ref{fig:partition}).  The difference $\delta
x_u(n_l\!=\!32)$$-$$\delta x_u(n_l\!=\!16)\simeq0.03-0.05$ nm from the
calculation of phase space volumes (at forces of order
$200-300\,\mathrm{pN}$) agrees with the difference
$x_u(n_l\!=\!32)$$-$$x_u(n_l\!=\!16)=0.026\,\mathrm{nm}$ measured from
pulling simulations.  This gives us confidence that for this model of
protein L the transition state is sharp and its volume does not change
appreciably under an applied force.

\textit{Discussion---} We have shown that an externally applied force
restricts a fluctuating protein's accessible phase space volume, which
increases the transition state displacement $x_u^{\mathrm{1D}}$ in the
equivalent one dimensional two-state model.  This contribution can be
appreciable because of the many degrees of freedom in a protein.
Larger proteins have a potentially larger $x_u^{\mathrm{1D}}$,
depending on which degrees of freedom couple to the applied force;
this will depend critically on the topology of the fold and the
direction in which it is pulled.  Most importantly, we predict that
$x_u^{\mathrm{1D}}$ should be greater for proteins unfolded through
longer attached linker strands.  This may have biological
significance; \textit{e.g.}  the long unfolded PEVK regions in titin
\cite{linke98} may play help modify the unfolding characteristics of
titin.  Finally, we note that many experiments have unfolded
concatamers of multiple domains, for convenience of attachment and to
generate larger statistics
\cite{rief99,best01,carrion-vazquez99,brockwell02,brockwell03,carrion-vazquez03,best03,williams03,brockwell05}.
We surmise that in all of these cases the dynamic contribution to
$x_u^{\mathrm{1D}}$ was significant, and also included a contribution
from already unfolded domains, which act as ``linkers'' for the last
few domains to unfold in a given pull.

\textit{Acknowledgements---} DKW acknowledges the Wellcome Trust for a
PhD studentship.  We thank D.~J. Brockwell, J. Clarke, T.
McLeish, and S. Radford for helpful discussions.


\begin{thebibliography}{10}

\bibitem{rief97}
M. Rief, M. Gautel, F. Oesterhelt, J.~M. Fernandez, and H.~E. Gaub, Science
  {\bf 276},  1109  (1997).

\bibitem{rief98a}
M. Rief, J.~M. Fernandez, and H.~E. Gaub, Phys. Rev. Lett. {\bf 81},  4764
  (1998).

\bibitem{west06}
D.~K. West, D.~J. Brockwell, P.~D. Olmsted, S.~E. Radford, and E. Paci,
  Biophys. J. {\bf 90},  287  (2006).

\bibitem{ortiz05}
V. Ortiz, S.~O. Nielsen, M.~L. Klein, and D.~E. Discher, J. Mol. Biol. {\bf
  349},  638  (2005).

\bibitem{bell78}
G.~I. Bell, Science {\bf 200},  618  (1978).

\bibitem{fernandez04}
J.~M. Fernandez and H. Li, Science {\bf 303},  1674  (2004).

\bibitem{schlierf04}
M. Schlierf, H. Li, and J.~M. Fernandez, Proc. Natl. Acad. Sci. USA {\bf 101},
  7299  (2004).

\bibitem{oberhauser01}
A.~F. Oberhauser, P.~K. Hansma, M. Carrion-Vazquez, and J.~M. Fernandez, Proc.
  Natl. Acad. Sci. USA {\bf 98},  468  (2001).

\bibitem{brockwell05}
D.~J. Brockwell, G.~S. Beddard, E. Paci, D.~K. West, P.~D. Olmsted, D.~A.
  Smith, and S.~E. Radford, Biophys. J. {\bf 89},  506  (2005).

\bibitem{brockwell02}
D.~J. Brockwell, G.~S. Beddard, J. Clarkson, R.~C. Zinober, A.~W. Blake, J.
  Trinick, P.~D. Olmsted, D.~A. Smith, and S.~E. Radford, Biophys. J. {\bf 83},
   458  (2002).

\bibitem{schlierf06}
M. Schlierf and M. Rief, Biophys. J. {\bf 90},  L33  (2006).

\bibitem{hummer03}
G. Hummer and A. Szabo, Biophys. J. {\bf 85},  5  (2003).

\bibitem{dudko2006}
O.~K. Dudko, G. Hummer, and A. Szabo, Phys. Rev. Lett. {\bf 96},  108101
  (2006).

\bibitem{Bartolo02}
D. Bartolo, I. Der\'enyi, and A. Ajdari, Phys. Rev. E {\bf 65},  051910
  (2002).

\bibitem{williams03}
P.~M. Williams, S.~B. Fowler, R.~B. Best, J. Toca-Herrera, K.~A. Scott, A.
  Steward, and J. Clarke, Nature {\bf 422},  446  (2003).

\bibitem{li04}
P.~C. Li and D.~E. Makarov, J. Chem. Phys. {\bf 121},  4826  (2004).

\bibitem{kirmizialtin05}
S. Kirmizialtin, L. Huang, and D.~E. Makarov, J Chem Phys {\bf 122},  234915
  (2005).

\bibitem{carrion-vazquez03}
M. Carrion-Vazquez, H. Li, H. Lu, P.~E. Marszalek, A.~F. Oberhauser, and J.~M.
  Fernandez, Nature Struct. Biol. {\bf 10},  738  (2003).

\bibitem{brockwell03}
D.~J. Brockwell, E. Paci, R.~C. Zinober, G.~S. Beddard, P.~D. Olmsted, D.~A.
  Smith, R.~N. Perham, and S.~E. Radford, Nature Struct. Biol. {\bf 10},  731
  (2003).

\bibitem{karanicolas02}
J. Karanicolas and C.~L. {Brooks III}, Prot. Sci. {\bf 11},  2351  (2002).

\bibitem{west06b}
D.~K. West, P.~D. Olmsted, and E. Paci, J. Chem. Phys. {\bf 124},  154909
  (2006).

\bibitem{haenggi90}
P. Hanggi, P. Talkner, and M. Borkovec, Rev. Mod. Phys. {\bf 62},  251  (1990).

\bibitem{evans97}
E. Evans and K. Ritchie, Biophys. J. {\bf 72},  1541  (1997).

\bibitem{oneill01}
J.~W. O'Neill, D.~E. Kim, D. Baker, and K.~Y. Zhang, Acta Crystallogr. D Biol.
  Crystallogr. {\bf 57},  480  (2001).

\bibitem{karanicolas03a}
J. Karanicolas and C.~L. {Brooks III}, J. Mol. Biol. {\bf 334},  309  (2003).

\bibitem{west06c}
D.~K. West, E. Paci, and P.~D. Olmsted, Phys. Rev. E. {\bf in preparation},
  (2006).

\bibitem{linke98}
W.~A. Linke, M. Ivemeyer, P. Mundel, M.~R. Stockmeier, and B. Kolmerer, Proc.
  Natl. Acad. Sci. USA {\bf 95},  8052  (1998).

\bibitem{rief99}
M. Rief, J. Pascual, M. Saraste, and H.~E. Gaub, J. Mol. Biol. {\bf 286},  553
  (1999).

\bibitem{best01}
R.~B. Best, B. Li, A. Steward, V. Daggett, and J. Clarke, Biophys. J. {\bf 81},
   2344  (2001).

\bibitem{carrion-vazquez99}
M. Carrion-Vasquez, A.~F. Oberhauser, S.~B. Fowler, P.~E. Marszalek, S.~E.
  Broedel, J. Clarke, and J.~M. Fernandez, Proc. Natl. Acad. Sci. USA {\bf 96},
   3694  (1999).

\bibitem{best03}
R.~B. Best, S. Fowler, J.~L. Toca-Herrera, A. Steward, E. Paci, and J. Clarke,
  J. Mol. Biol. {\bf 330},  867  (2003).

\end{thebibliography}
\end{document}